\documentclass[twocolumn,showpacs,preprintnumbers,amsmath,amssymb,superscriptaddress]{revtex4}
\usepackage{graphicx}
\usepackage{dcolumn}
\usepackage{bm}
\usepackage{epsfig}
\usepackage{CJK}
\usepackage{bm}

\begin{document}

\title{Manipulating Edge Majorana Fermions in Vortex State}

\author{Qi-Feng Liang}
\affiliation{International Center for Materials Nanoarchitectornics (WPI-MANA)
National Institute for Materials Science, Tsukuba 305-0044, Japan}
\affiliation{Department of Physics, Shaoxing University, Shaoxing 312000, China}
\author{Zhi Wang}
\author{Xiao Hu}
\affiliation{International Center for Materials Nanoarchitectornics (WPI-MANA)
National Institute for Materials Science, Tsukuba 305-0044, Japan}

\date{\today}

\begin{abstract}
A vortex in a model spinless $p_x+i p_y$ superconductor induces two
Majorana fermions (MFs), one in the core and the other at the sample
edge. In the present work, we show that edge MF can be generated,
fused, transported, and braided easily by tuning gate voltages at
point-like constriction junctions. Solving the time-dependent
Bogoliubov-de Gennes equation, we demonstrate that the braiding of
edge MFs obeys the non-Abelian statistics. The present setup is
therefore a promising implementation for topological
quantum computation, and has the advantage of easy manipulation and
simple device structure.
\end{abstract}

\maketitle

Topological quantum computation is attracting considerable
interests due to the unique feature of fault tolerance, where
quantum information is stored non-locally and robust against
decoherence caused by interaction with environment \cite{TQC_RMP,
Kitaev}. Ground
state degeneracy and rotations within the degenerate subspace with
unitary non-Abelian transformation are the two important
ingredients for its implementation.
Systems with Majorana fermions (MFs)
\cite{mooreread91,WXG,ReadGreen,Ivanov,Kitaev01,DasSarmaPRL05,superfluid,Gurarie,
FuKanePRL08,MFreturn,
TKNg,YouJQFNori,SauDasSarmaPRL10,beenakker,Nagaosa,LutchynDasSarma,OregvonOppen10,FisherNPhys,ShenSQ,TPSC,Fwave,HQV}, particles equivalent to their
antiparticles, as zero-energy excitations are promising, since they
are half of conventional fermions and can form non-local qubits.
Superconducting state provides a
natural host for these zero-energy MFs where
Bogoliubov quasiparticles are composed by both electrons and holes.

It was illustrated that a vortex in spinless $p_x+i p_y$
superconductor can accommodate a zero-energy MF at its core
\cite{ReadGreen}, and the braiding of vortices obeys the non-Abelian
statistics \cite{Ivanov}. A hetero structure of $s$-wave
superconductor (S) and topological insulator (TI) \cite{TI1,TI2,TI3}
was then proposed to behave similarly to the spinless $p$-wave
superconductor due to strong spin-orbit coupling, and circuits of
S/TI/S junctions can achieve creation, manipulation and fusion of
MFs by tuning the superconductivity phases \cite{FuKanePRL08}. Very
recently, it was demonstrated that a spin-orbit coupling
semiconductor (SM) in proximity to a ferromagnetic insulator (FI)
plays a similar role as TI, and thus a S/SM/FI hetero structure can
be a generic platform to provide topological phase
\cite{SauDasSarmaPRL10}. It was also illustrated clearly by a toy
model that a one-dimensional (1D) spinless superconductor carries
two zero-energy MFs at its two ends \cite{Kitaev01}, which can be
realized by a 1D spin-orbit coupling semiconductor under a magnetic
field and in proximity to an $s$-wave superconductor
\cite{LutchynDasSarma,OregvonOppen10}. The non-Abelian braiding of
end MFs has been demonstrated very recently in a network of these
nanowires with voltage applications \cite{FisherNPhys}.

\begin{figure}[t]
\psfig{figure=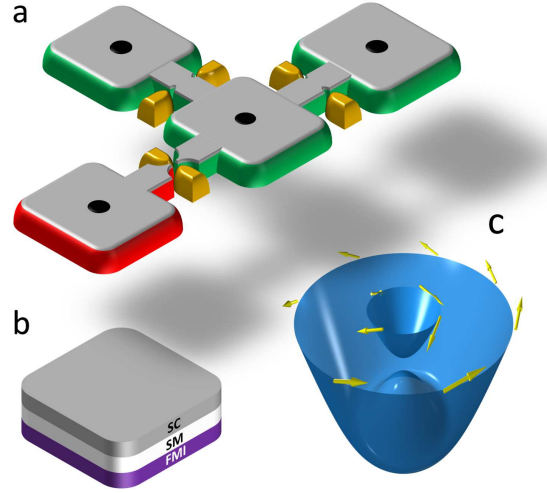,width=8cm} \caption{(a) Schematic device
setup for generating and braiding of edge MFs by tuning gate voltage
at point-like constriction junctions. (b) Each of the four finite samples
(called bricks)
consists of a hetero structure of $s$-wave superconductor with a
vortex, spin-orbit coupling semiconductor and a ferromagnetic
insulator. (c) Tuning chemical potential to the gap between the two
bands, a superconducting gap is opened at the outside one.}
\end{figure}

In the present work, we notice that the edge MF realized in a finite
sample with a superconducting vortex is very useful. Since MFs
always come as pairs, the zero-energy MF bounded at the vortex core
has its counterpart, which appears at the sample edge (see
Fig.~1(a)). The edge MFs have important advantages in manipulation,
namely they can be created, fused, transported and braided easily by
applying gate voltages on constriction junctions between finite
samples as revealed in the present work. In contrast, motion of
core MFs should always be accompanied by vortices, which is not an
easy task practically. Comparing with previous proposals based on
control of superconductivity phases in circuit of S/TI/S junctions
\cite{FuKanePRL08} and manipulation of end MFs in network of 1D
nanowires \cite{FisherNPhys}, gate voltages are applied at the
point-like constriction junctions in the present setup, which makes
the device structure and operations simpler.  As a drawback of the
present device, the energy gap between edge zero-energy MFs and
finite-energy excitations is smaller compared with the core
counterparts, thus requiring manipulation at lower temperature. It
is expected, however, that this difficulty can be overcome
relatively easier by using superconductor with large energy gap.

Explicitly we consider a finite sample of the S/SM/FI hetero
structure with a superconducting vortex at the sample center
\cite{LutchynDasSarma, SauDasSarmaPRL10}, which we call a brick.
Bricks are connected by constriction junctions, where gate voltages
can be tuned to connect and disconnect the constriction junctions
adiabatically. We show that edge MFs can be created, fused,
transported, and braided with sequences of switching on and off the
gate voltages in the system. Solving the time-dependent
Bogoliubov-de Gennes (BdG) equation, we monitor the time evolution
of the wave functions, and demonstrate clearly that braiding of edge
MFs obeys the non-Abelian statistics.

\vspace{5mm}
\noindent\textbf{Bogoliubov-de Gennes Hamiltonian}

\noindent Our elementary system is schematically depicted in
Fig.~1(b). The Hamiltonian of a spin-orbit coupling semiconductor in
proximity to a ferromagnetic insulator is given by

\begin{equation}
H_0= \int d\vec{r} \tilde{c}^{\dagger}(\vec{r})
  \left[\frac{\overrightarrow {p}^2}{2m^*} -\mu
         + \alpha_R (\overrightarrow {\sigma}\times \overrightarrow{p} ) \cdot \hat{z}
         + V_z \hat{\sigma}_z \right] \tilde{c}(\vec{r})
\end{equation}
with $m^*$, $\mu$, $\alpha_R$ and $V_z$ being the effective electron
mass, chemical potential, strength of the Rashba spin-orbit
coupling, and Zeeman field, respectively, and
$\vec{\sigma}=(\hat{\sigma}_x,\hat{\sigma}_y,\hat{\sigma}_z)$
the Pauli matrices and $\tilde c =(\hat c_{\uparrow} \hspace{1mm} \hat c_{\downarrow})^{\rm T}$
the electron annihilation operators.
The proximity effect from the $s$-wave superconductor is described by
\begin{eqnarray}
H_{\rm sc}=\int d
\vec{r}[\Delta(\vec{r})\hat{c}_{\uparrow}^{\dag}(\vec{r})\hat{c}_{\downarrow}^{\dag}(\vec{r})+h.c.],
\end{eqnarray}
where  $\Delta(\vec{r})$ is the effective $s$-wave pairing
potential. The total Hamitonian  is $H_{\rm tot}=H_0+H_{\rm sc}$.
For chemical potential $\sqrt{\Delta^2+\mu^2}<V_z$, the system
behaves effectively as a 2D spinless $p_x+i p_y$ superconductor with
the dispersion and spin configuration given in Fig.~1(c)
\cite{ReadGreen,DasSarmaPRB10,LM_ZCW,Gurarie, Tewari_Index}.

The quasiparticle excitations in the system are described by the
following Bogoliubov-de-Gennes (BdG) equation
\begin{equation}
\left(
\begin{array}{cc}
H_0& \Delta\\
\Delta^{\dag} & -\hat\sigma_y H_0^{*}\hat\sigma_y\\
\end{array}
\right)\Psi(\vec{r})=E\Psi(\vec{r}),
\end{equation}
where the Nambu spinor notation
$\Psi(\vec{r})=\left[u_{\uparrow}(\vec{r}),u_{\downarrow}(\vec{r}),
v_{\downarrow}(\vec{r}),-v_{\uparrow}(\vec{r})\right]$ is adopted.
Since $H_0$ is Hermitian, any eigen vector $\Psi$ of Eq.(3) with
energy $E$ has its twin $\hat\sigma_y \hat\tau_y \Psi^*$ with energy $-E$,
where $\hat\tau_y$ is the Pauli matrix in Nambu spinor space.
Because the BdG equation is defined by an even dimensional Hamiltonian, zero-energy eigen
modes, if any, should appear in pairs. By recombining these
zero-energy eigen wave functions, one can always achieve $\Psi=
\hat\sigma_y \hat\tau_y \Psi^*$, for which the quasiparticle operator defined by

\begin{equation}
\hat{\gamma}^{\dag}=\int
d\vec{r}\displaystyle\sum_{\sigma}u_{\sigma}
(\vec{r})\hat{c}_{\sigma}^{\dag}(\vec{r})+v_{\sigma}(\vec{r})\hat{c}_{\sigma}(\vec{r})
\end{equation}
satisfies the relation $\hat{\gamma}^{\dag}=\hat{\gamma}$.
Therefore, the zero-energy excitations of the system are actually
pairs of MFs.

It has been shown \cite{DasSarmaPRB10, LM_ZCW, Gurarie} that the
system Eq.~(3) with a vortex in the superconductor accommodates a MF
at the vortex core. For the case that there is only one vortex in
the system, its twin should appear at the edge of the system
\cite{Tewari_Index,PNAS}. While the core MF has been investigated in
many previous studies, little attention has been paid on the edge MF
(see \cite{FuKanePRL08}), which is the main focus of the present
work.

In order to explore the edge MF, we resort to numerical analysis.
For this purpose, we derive the tight-binding version of $H_0$ on a
square grid,

\begin{eqnarray}
\tilde{H}_0=&&\displaystyle-t_0\sum_{\bold{i},\bold{j},\sigma}
\hat{c}_{\bold{i}\sigma}^{\dag}\hat{c}_{\bold{j}\sigma}
- \mu \sum_{\bold{i},\sigma} \hat{c}_{\bold{i}\sigma}^{\dag}\hat{c}_{\bold{i}\sigma}+\sum_{\bold{i}}V_z (\hat{c}_{\bold{i}\uparrow}^{\dag}\hat{c}_{\bold{i}\uparrow}-\hat{c}_{\bold{i}\downarrow}^{\dag}\hat{c}_{\bold{i}\downarrow}) \nonumber\\
&&+\displaystyle i t_{\alpha} \sum_{\bold{i},\bold{\delta}}\left[\hat{c}_{\bold{i}+\bold{\delta}_x}^{\dag}\hat{\sigma}_y\hat{c}_{\bold{i}}-
\hat{c}^{\dag}_{\bold{i}+\bold{\delta}_y}\hat{\sigma}_x\hat{c}_{\bold{i}}+h.c.\right],\\\nonumber
\end{eqnarray}
where both spin-reserved
hopping $t_0=\hbar^2/2m^*a^2$ and spin-flipped hopping
$t_{\alpha}=\alpha_R\hbar/2a$ are between nearest neighbors with $a$
the grid spacing, and $\mu$ is measured from band bottom.

The tight-binding BdG equation has the same form of Eq.(3), and
becomes 4$N$ coupled linear equations with $N$ being the total
number of grid sites. The energy spectrum and all wave functions of
excitations including those of MFs can be obtained by diagonalizing
the $4N\times 4N$ matrix. In the present study, the square sample is divided into a grid of
$100\times 100$ sites. Although the size of the Hamiltonian matrix
can be large typically in the order of $10^5$, in most cases only
several states near $E=0$ are relevant, for which a powerful
numerical technique is available \cite{JADAMILU}.

\begin{figure}[t]
\psfig{figure=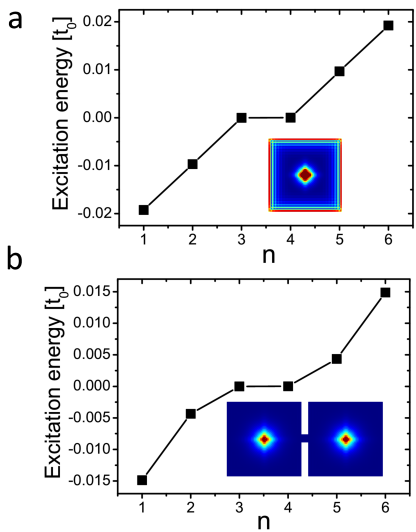,width=6cm} \caption{Energy spectrum of
several lowest excitations and distribution of wave-function norms
of zero-energy MFs, (a) for one brick with two MFs, one at vortex core and the other
at brick edge, and (b) for two connected bricks with two core MFs.
Results are for $\Delta_0=0.5t_0$, $V_z=0.8t_0$, $\mu=0$ and
$t_\alpha=0.9 t_0$, with $100 \times100 $ sites for the brick
and $10 \times 10$ sites for the constriction junction. }
\end{figure}

\begin{figure}[t]
\psfig{figure=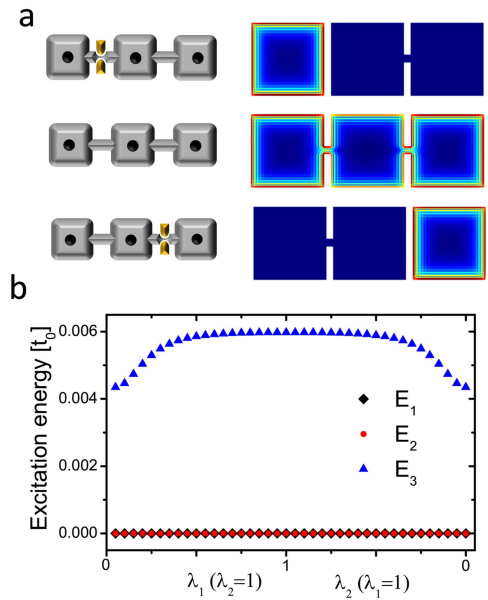,width=7cm} \caption{(a) Diagram for transport
of edge MF prepared at the left brick to the right brick and
distributions of calculated wave-function norms. (b) Evolution of
the energy spectrum, where the two zero-energy excitations are
associated with three core and one edge MFs, when the left junction
is turned on adiabatically parameterized by $\lambda_1$, and then
the right junction is turned off adiabatically parameterized by
$\lambda_2$.}
\end{figure}

\vspace{5mm}
\noindent\textbf{Edge Majorana Fermions}

\noindent We first study a square sample (see Fig.~1(b))
 with a vortex at the center $\Delta(\vec{r})=
\Delta_0[1-e^{-(r/s)^2}]e^{i\theta}$ with $s=4a$, namely the
brick. As shown in the energy spectrum in Fig.~2(a), we found a pair
of zero-energy excitations in the system. By recombination, we get
two wave functions, one at the vortex core and the other on the
sample edge with their norms plotted in the inset of Fig.~2(a), both satisfying
the generic relation for MF. We define them as core MF and
edge MF.  This treatment of wave functions is understood for all
discussions through the present work.

Next we study two bricks connected by a constriction junction. As
shown in Fig.~2(b), we find again two zero-energy MFs, each bounded
by one of the two vortices. There is no edge MF this time (see inset of
Fig.~2(b)), since the
two edge MFs in the two bricks meet each other at the constriction
junction and fuse into a Bogoliubov quasiparticle with finite
energy.

We proceed to investigate a system of three bricks, and introduce
dynamic process of switching on and off constriction junctions. This
is implemented by tuning the hopping integrals at the constriction
junctions among bricks, which can be realized practically by
adjusting gate voltage. At the initial state, the left brick is
isolated while the other two are connected. There are three core and
one edge MFs as revealed above. As confirmed by calculation, the
core MFs remain unchanged (in an exponential accuracy) during the
switching process, we omit them in the following discussions for
simplicity. The constriction junction between the left and central
bricks is then turned on adiabatically. As displayed in Fig.~3(a),
it is interesting to observe that, the edge MF on the left brick
spreads its wave function to the central and right bricks, since now
there is only one unified edge of the three bricks. We emphasize
that this is one of the clearest manifestations of the mobility of
the edge state.

\begin{figure*}[t]
\psfig{figure=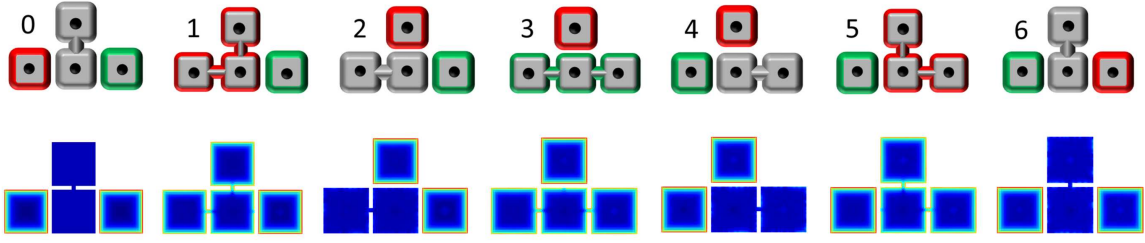,width=16cm} \caption{Diagram for exchanging
two edge MFs (red and green for eye-guide) and distribution of
wave-function norms. }
\end{figure*}

We then turn off the constriction junction between the central brick
and the right one adiabatically. As shown in Fig.~3(a), the wave
function of edge MF now shrinks to the right brick. As the result of
these two switchings, the edge MF initially at the left brick is
transported to the right brick.

During the whole process, we monitor the energy spectrum and make
sure that the gap between the zero-energy MFs and excitations of
finite energy remains open (see Fig.~3(b)), which protects the
topological phase of the system. Compared with previous
prescriptions \cite{FuKanePRL08,FisherNPhys}, the present manipulations
on edge MFs are performed by applications of point-like gate
voltages, which makes the device simpler.

\vspace{5mm}
\noindent\textbf{Braiding with Non-Abelian Statistics}

\noindent Based on the above results, we can exchange edge MFs pairwisely. The
simplest structure for this purpose consists of four bricks, as
depicted in Fig.~1(a), and the diagram of exchange process of two edge MFs
are demonstrated in
Fig.~4. As the initial state (step-0), there are two edge MFs
located at the left and right bricks, whereas the central and top
bricks are connected (thus no edge MFs). Following the process for
transportation revealed above, we first transport the red edge MF to
the top brick (step-1, -2). We then transport the green edge MF from
the right brick to left brick (step-3, -4). Finally, we bring the
red edge MF tentatively stored at the top brick to the right brick
(step-5, -6). After these operations, the setup of the system, or
the Hamiltonian, returns to the initial one, but leaving the two
edge MFs exchanged.

Now let us reveal the impact of this exchange to the wave
function of system. For this purpose, we evaluate the time evolution of wave
function upon adiabatic switchings by using the time-dependent BdG
(TDBdG) method \cite{tdbdg_1,tdbdg_2,tdbdg_3}. We first derive the wave functions of edge MFs at
the left and right bricks by diagonalizing the Hamitonian Eq.(3),
with which the initial wave function for the total system is given
by $|\Psi(t=0)\rangle=|\phi_L\rangle+|\phi_R\rangle$.
Upon on the switching process illustrated in
Fig.~4, the wave function evolves with time according to the TDBdG
equation:

\begin{equation}
   i\hbar\frac{d|\Psi(t)\rangle}{dt}=H_{\rm tot}(t)|\Psi(t)\rangle.
\end{equation}
The evolution of wave function can be monitored by its projections
to the initial wave functions of the two edge MFs,
$O_L\equiv\langle\phi_L|\Psi(t)\rangle$ and
$O_R\equiv\langle\phi_R|\Psi(t)\rangle$. At the initial state (step-0), one has
$O_L=O_R=+1$ by definition. As shown in Fig.~5, the order parameter
$O_L$ drops gradually since the wave function of the red edge MF
spreads to the three bricks when the left junction is turned on
(step-1 in Fig.~4); it drops to zero when the red edge MF is totally
transported to the top brick (step-2). When the right junction is
then turned on and thus the green edge MF spreads its wave function
to the central and left bricks (step-3), $O_L$ becomes finite, and
most intriguingly it acquires a minus sign. It achieves $O_L=-1$ when
the left junction is switched off (step-4), and remains its value
for the final two steps.

\begin{figure}[t]
\psfig{figure=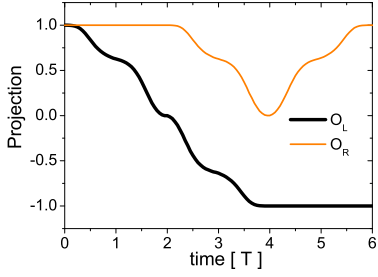,width=6cm} \caption{Time evolution of the
wave function of edge MFs for the exchange process in Fig.~4, in
terms of its projections $O_L$ and $O_R$ to the two edge MFs
prepared at the initial stage. The time for a transport process
is taken as $T=40000 \hbar/t_0$. }
\end{figure}

The evolution of the order parameter $O_R$ up to step-4 is
straightforward. When the red edge MF stored tentatively at
the top brick spreads to the right brick (step-5 and -6), $O_R$
becomes positive and then achieves $O_R=+1$ by the time the whole
process is finished.

Therefore, during the process of exchanging the two edge MFs illustrated in Fig.~4,
the wave function of the system evolves as

\begin{equation}
|\Psi(t=0)\rangle=|\phi_L\rangle+|\phi_R\rangle \Rightarrow
|\Psi(t=6T)\rangle=-|\phi_L\rangle+|\phi_R\rangle,
\end{equation}
or in terms of operators,

\begin{equation}
\hat{\gamma_L}\rightarrow\hat{\gamma_R}, \hspace{10mm}
\hat{\gamma_R}\rightarrow-\hat{\gamma_L},
\end{equation}
which can be presented by a unitary transformation
$U_{LR}=e^{\frac{\pi}{4}\hat{\gamma}_R\hat{\gamma}_L}$, the same
form observed first for core MFs \cite{Ivanov} (see also
\cite{FisherNPhys}).

The above exchange rule of edge MFs can be understood in the
following way. Let us start from the MF transport in Fig.~3. An
effective Hamiltonian for the low-energy physics of
the adiabatic transport is given by

\begin{equation}
 \hat{H}_{\rm eff}(t) = i\lambda_1(t) \Gamma_{LC}\hat{\gamma}_L \hat{\gamma}_C
                   + i\lambda_2(t) \Gamma_{CR}\hat{\gamma}_C \hat{\gamma}_R,
\end{equation}
where $\Gamma_{ij}=-\Gamma_{ji}$ denotes interaction between MFs, and
for edge MF transport $\lambda_1(t)$ increases from 0 to 1
from $t=0$ to $t=T$ while $\lambda_2$ remains unity, and from $t=T$
to $t=2T$ $\lambda_2(t)$ decreases from 1 to 0 while $\lambda_1$
remains unity. A zero-energy edge MF can be composed from the three
MF creation operators at any moment provided the topological phase
is protected by the adiabatic processes

\begin{equation}
 \hat{\gamma}(t) = \frac{\hat{\gamma}_L\lambda_2(t)}{\sqrt{[\lambda_1(t)\frac{\Gamma_{LC}}{\Gamma_{CR}}]^2+\lambda_2^2(t)}}
+\frac{\hat{\gamma}_R\lambda_1(t)\Gamma_{LC}/\Gamma_{CR}}{\sqrt{[\lambda_1(t)\frac{\Gamma_{LC}}{\Gamma_{CR}}]^2+\lambda_2^2(t)}},
\end{equation}
with $\hat{\gamma}(t=0)=\hat{\gamma}_L$ and
$\hat{\gamma}(t=2T)={\rm sgn}(\Gamma_{LC}/\Gamma_{CR})\hat{\gamma}_R$.
An interaction between $\gamma_L$ and $\gamma_R$ can be included, but it is
easy to see that the result of transport remains the same.

The exchanging process which consists of three transport processes with
two edge MFs as shown in Fig.~4 is then described in the following way:
$\hat{\gamma}_1(t=2T)={\rm sgn}(\Gamma_{LC}/\Gamma_{CT})\hat{\gamma}_T$ and
$\hat{\gamma}_2(t=2T)=\hat{\gamma}_R$,
$\hat{\gamma}_1(t=4T)=\hat{\gamma}_1(t=2T)$ and
$\hat{\gamma}_2(t=4T)={\rm sgn}(\Gamma_{RC}/\Gamma_{CL})\hat{\gamma}_L$,
$\hat{\gamma}_1(t=6T)= {\rm sgn}(\Gamma_{TC}/\Gamma_{CR}) {\rm sgn}(\Gamma_{LC}/\Gamma_{CT})\hat{\gamma}_R =-{\rm
sgn}(\Gamma_{LC}/\Gamma_{CR})\hat{\gamma}_R$ and
$\hat{\gamma}_2(t=6T)=\hat{\gamma}_2(t=4T)$. It is then clear that
after the braiding, the two MFs at the left and right bricks acquire
opposite signs, as revealed in Fig.~5.

Now we are ready to exhibit that the braiding of edge MFs obeys the
non-Abelian statistics. As illustrated in Fig.~6 with five bricks
and three edge MFs, starting from the same initial state, the two
processes of different orders of braiding generate two different
final states, which corresponds directly to the inequality
$U_{LT}U_{TR} \neq U_{TR}U_{LT}$, namely the non-Abelian feature of
braiding.

\begin{figure}[t]
\psfig{figure=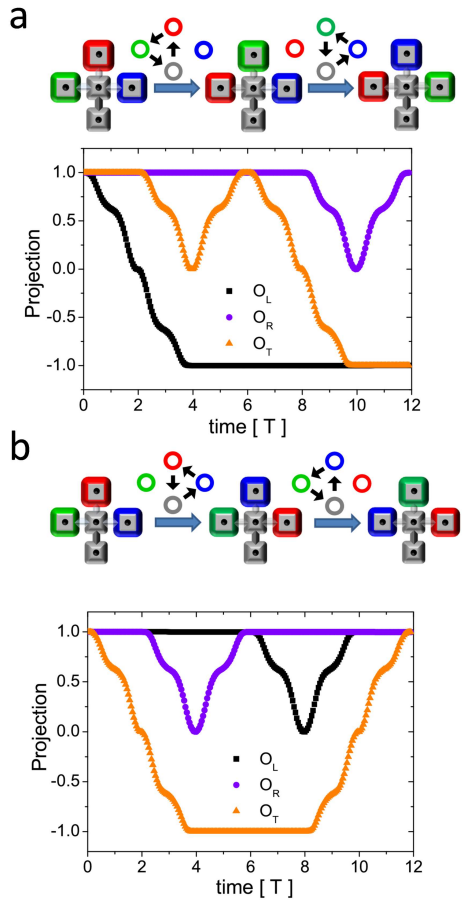,width=7cm} \caption{Diagrams for two braiding
processes of three edge MFs (red, green and blue for eye-guide) and
time evolution of the wave function of edge MFs, in terms of its
projections $O_L$, $O_R$ and $O_T$ to the three edge MFs prepared at
the initial stage. }
\end{figure}

\vspace{5mm}
\noindent\textbf{Discussions}

\noindent In an implementation of topological quantum computation
based on the zero-energy MF modes, the working temperature is
limited by the energy gap to the lowest excitation. The excitation
gap is smaller in the present setup based on edge MFs than an
implementation using core MFs simply because the edge MFs are
distributed over the sample edge. Quantitatively, the excitation gap
is about $0.01\Delta_0$ for parameters $\Delta_0=0.5t_0$,
$V_z=0.8t_0$ and $\alpha_R=0.9t_0$. For a superconductivity gap
$\Delta_0\sim 1$meV \cite{DasSarmaPRB10,Nb}, the corresponding
temperature of the excitation gap will be $\sim 100$mK, which is not
hard for low-temperature technologies in these days.

While not appearing apparently in the criterion for chemical
potential $V_z>\sqrt{\Delta^2+\mu^2}$ supporting the topological
phase, the spin-orbit coupling $\alpha_R$ does play an important
role in determining the magnitude of effective $p$-wave pairing and
thus the excitation gap. Besides InAs \cite{SauDasSarmaPRL10}, the
layered polar semiconductor BiTeI with giant spin-orbit coupling can
be a promising material for our device \cite{GiantSOC}. The Zeeman
splitting $V_z$ of order of several meV can be realized in the thin
film of strong ferromagnetic insulator \cite{EuO} according to a
recent work \cite{DasSarmaPRB10}.

The present setup has good scalability, and an
array of the units in Fig.~1(a) supports a ground state wave
function consisted of a linear combination of the many-body Majorana
edge states. The ground state can be rotated in the degenerate
subspace by exchanging pairwisely the Majorana edge states, which is
given by a multidimensional unitary matrix representation of the 2D
braid group, and governed by the non-Abelian statistics.

To conclude, we have demonstrated that the edge MFs induced by
vortices in topological superconductor can be manipulated easily by
application of gate voltages at point-like constriction junctions.
Adiabatic braidings of edge MFs have been simulated by
TDBdG method, and the non-Ableian
statistics is proved. The present proposal therefore provides a
promising way for implementing quantum topological computation
based on zero-energy MFs, and has the advantage of easy operation
and simple device structure.

\vspace{5mm}
\noindent\textbf{Methods}

\noindent In this work, we use TDBdG method to simulate the
adiabatic braiding of edge MFs upon switching on and off the
constriction junctions. The main task of the simulation is to solve
the TDBdG equation (6). The formal solution of the equation is
$|\Psi(t)\rangle = \exp(-i\hbar \int H(t)dt)|\Psi(0)\rangle$. To solve it
numerically, we first divide the total simulation time into small
steps of $\Delta t$, in which the Hamiltonian can be considered as
time independent, and thus $|\Psi(t+\Delta t)\rangle \simeq
\exp(-i\hbar H(t)\Delta t)|\Psi(t)\rangle$. The exponential can be
expanded by the Chebyshev polynomials $T_n(H)$
\begin{equation}
 e^{-i\hbar H(t)\Delta t}=\displaystyle\sum_{m}^{\infty}c_m(\Delta t)T_m(H).
\end{equation}
Because the coefficients $c_m$ decrease exponentially \cite{tdbdg_1,tdbdg_2},
only a finite number ($M_{\rm
max}$) of terms in the above expansion are needed for sufficient
accuracy. Using the recursive relations of the Chebyshev polynomials
$T_m(H)=2HT_{m-1}(H)-T_{m-2}(H)$, the summation in Eq.(11) can be
performed by $M_{\rm max}$ times matrix-multiplication. For a sparse
matrix $H$ in our system, the computation time is of $O(N)$
\cite{tdbdg_1}.

The time-dependent part of the Hamiltonian is the hopping integrals
at the sites on constriction junctions between bricks. In the
present study, we use a parameter $0\le \lambda(t)\le 1$ to scale
the hopping integrals between zero to $t_0$. In order to simulate
adiabatic processes, the function $\lambda(t)$ should vary slowly
with $t$, and thus the total evolution time $T$ is taken
sufficiently long as compared with the inverse of the lowest
excitation energy. Furthermore, we use $\lambda(t)=(t/T)^2$ rather
than a linear function of $t$, which suppresses efficiently excitations
from zero-energy MFs to states with finite energy.

\vspace{5mm}
\noindent\textbf{Acknowledgements}

\noindent This work was supported by WPI Initiative on Materials
Nanoarchitectonics, MEXT of Japan, and Grants-in-Aid for Scientific
Research (No.22540377), JSPS, and partially by CREST, JST. Q.F.L. is
also supported by NSFC under grants 10904092.


\begin{thebibliography}{99}
\bibitem{Kitaev} Kitaev, A. Y. Fault-tolerant quantum computation by anyons. Ann. Phys. \textbf{303}, 2 (2003).
\bibitem{TQC_RMP} Nayak, C., Simon, S. H., Stern, A., Freedman, M. \& Das Sarma, S. Non-Abelian anyons
and topological quantum computation. Rev. Mod. Phys. \textbf{80}, 1083 (2008).

\bibitem{mooreread91} Moore, G. \& Read, N. Nonabelions in the
fractional quantum hall effect. Nuc. Phys. B \textbf{360}, 362 (1991).

\bibitem{WXG} Wen, X. G. Non-Abelian topological order in $\nu$=1/2 quantum hall
state. Phys. Rev. Lett. \textbf{70}, 355–358 (1993).

\bibitem{ReadGreen} Read, N. \& Green, D. Paired states of fermions in two dimensions with breaking of parity
and time-reversal symmetries and the fractional quantum hall effect. Phys. Rev. B \textbf{61}, 10267 (2000).

\bibitem{Ivanov} Ivanov, D. A. Non-Abelian statistics of half-quantum vortices in $p$-wave
superconductors. Phys. Rev. Lett. \textbf{86}, 268 (2001).
\bibitem{Kitaev01} Kitaev, A. Y. Unpaired majorana fermions in quantum wires. Phys. -Usp. \textbf{44}, 131 (2001).
\bibitem{DasSarmaPRL05} Das Sarma, S., Freedman, M. \& Nayak, C. Topologically protected qubits from a possible
non-Abelian fractional quantum hall state. Phys. Rev. Lett. \textbf{94}, 166802 (2005).

\bibitem{superfluid} Tewari, S., Das Sarma, S., Nayak, C., Zhang,
C. -W. \& Zoller, P. Quamtum computation using vortices and majorana
zero modes of a $p_x+ip_y$ superfluid of fermionic cold atoms. Phys. Rev. Lett \textbf{98}, 010506 (2007).

\bibitem{Gurarie} Gurarie, V. \& Radzihovsky, L. Zero modes of two-dimensional chrial $p$-wave superconductors.
Phys. Rev. B \textbf{75}, 212509 (2007).

\bibitem{FuKanePRL08} Fu, L. \& Kane, C. L. Superconducting proximity effect and majorana fermions at
the surface of a topological insulator. Phys. Rev. Lett.
\textbf{100}, 096407 (2008).

\bibitem{MFreturn} Wilczek, F. Majorana returns. Nat. Phys.
\textbf{5}, 614(2009).

\bibitem{TKNg} Law, K. T., Lee, P. A. \& Ng, T. K. Majorana fermion
induced resonant andreev reflection. Phys. Rev. Lett.
\textbf{103}, 237001 (2009).

\bibitem{YouJQFNori} You, J. Q., Shi, X. F., Hu, X. D. \& Nori, F. Quantum emulation of a spin system with topological protected
ground states using superconducting quantum circuits. Phys. Rev. B \textbf{81}, 014505 (2010).

\bibitem{SauDasSarmaPRL10} Sau, J. D., Lutchyn, R. M., Tewari, S. \& Das Sarma, S. Generic new platform for topological
quantum computation using semiconductor heterostructures. Phys. Rev.
Lett. \textbf{104}, 040502 (2010).

\bibitem{beenakker} Wimmer, M., Akhmerov, A. R., Medvedyeva, M. V., Tworzydlo, J. \& Beenakker, C. W. J.
Majorana bound states without vortices in topological superconductors with electrostatic defects.
Phys. Rev. Lett. \textbf{105}, 046803 (2010).
\bibitem{Nagaosa} Linder, J., Tanaka, Y., Yokoyama, T., Sudbø, A. \& Nagaosa, N. Unconventional
Superconductivity on a Topological Insulator. Phys. Rev. Lett. \textbf{104}, 067001 (2010).
\bibitem{LutchynDasSarma} Lutchyn, R. M., Sau, J. D. \& Das Sarma, S.
Majorana fermions and a topological phase transition in
semiconductor-superconductor heterostructures. Phy. Rev. Lett.
\textbf{105}, 077001 (2010)

\bibitem{OregvonOppen10} Oreg, Y., Rafael, G. \& von Oppen, F.
Helical liguids and majorana bound states in quantum wires. Phys.
Rev. Lett. \textbf{105}, 177002 (2010).

\bibitem{FisherNPhys} Alicea, J., Oreg, Y., Refael, G., von Oppen, F. \& Fisher, M. P. A. Non-Abelian statistics and topological
quantum information processing in 1D wire networks. Nat. Phys.
\textbf{7}, 412 (2011).

\bibitem{ShenSQ} Zhou, B. \& Shen, S. Q. Crossover from majorana
edge- to end-state in quasi-one-dimensional p-wave superconductors.
Phys. Rev. B \textbf{84}, 054532 (2011).


\bibitem{TPSC} Hosur, P., Ghaemi, P., Mong, R. S. K. \& Vishwanath, A. Majorana modes at the ends of superconductor vortices in doped topological
insulator. Phys. Rev. Lett. \textbf{107}, 097001 (2011).

\bibitem{Fwave} Mao, L., Shi, J. R., Niu, Q. \& Zhang, C. W. Superconducting phase with a chiral f-Wave
 pairing symmetry and majorana fermions induced in a Hole-Doped Semiconductor
Phys. Rev. Lett. \textbf{106}, 157003 (2011).

\bibitem{HQV} Jang, J., Ferguson, D. G., Vakaryuk, V.,
Budakian, R., Chung, S. B., Goldbart, P. M. \& Maeno, Y. Observation of half-height magnetization steps in $Sr_2RuO_4$.
 Science. \textbf{331}, 186 (2011).


\bibitem{TI1} Hasan, M. Z. \& Kane, C. L. colloquium:topological
insulators. Rev. Mod. Phys. \textbf{82}, 3045 (2010).

\bibitem{TI2} Yu, R., Zhang, W., Zhang, H. J., Zhang, S. C., Dai, X. \& Fang, Z.
Quantized anomalous hall effect in magnetic topological insulators. Science \textbf{329}, 61 (2010).


\bibitem{TI3} Qi, X. L. \& Zhang, S. C. Topological insulators and
superconductors. Rev. Mod. Phys. \text{83}, 1057 (2011).


\bibitem{Tewari_Index} Tewari, S., Das Sarma, S. \& Lee, D. -H., Index
theorem for the zero modes of majorana fermion vortices in chiral
$p$-wave superconductors. Phys. Rev. Lett. \textbf{99}, 037001 (2007).

\bibitem{DasSarmaPRB10} Sau, J. D., Tewari, S., Lutchyn, R. M., Stanescu, T. D. \& Das Sarma, S.
Non-Abelian quantum order in spin-orbit-coupled semiconductors:
search for topological majorana particles in solid-state systems.
Phys. Rev. B \textbf{82}, 214509 (2010).

\bibitem{LM_ZCW} Mao, L \& Zhang, C. W. Robustness of Majorana modes and minigaps in a spin-orbit-coupled
semiconductor-superconductor hererostructure. Phys. Rev. B \textbf{82}, 174506 (2010).

\bibitem{PNAS} Grofeld, E. \& Stern, A. Observing majorana bound states of josephson vortices in topological
superconductors. Proc. Nat. Acad. Sci. \textbf{108}, 11810 (2011).

\bibitem{JADAMILU} Bollh\"{o}fer, M. \& Notay, Y. JADAMILU:a software code for computing
selected eigenvalues of large sparse symmetric matrices. Comp. Phys. Com. \textbf{177}, 951 (2007).

\bibitem{tdbdg_1} Roche, S. Quantum transport by means of O(N) real-space methods. Phys. Rev. B \textbf{59}, 2284 (1999).
\bibitem{tdbdg_2} Loh, Y. L., Taraskin, S. N. \& Elliott, S. R. Fast time-evolution method
for dynamical system. Phys. Rev. Lett. \textbf{84}, 2290 (2000).
\bibitem{tdbdg_3} Wei{\ss}e, A., Wellein, G., Alveman, A. \& Fehske, H.
The Kernel polynomials method. Rev. Mod. Phys. \textbf{78}, 275 (2006).


\bibitem{Nb} Chrestin, A., Matsuyama, T. \& Merkt, U. Evidence for a
proximity-induced energy gap in Nb/InAs/Nb junctions. Phys. Rev. B \textbf{55} ,8457 (1997).


\bibitem{GiantSOC} Ishizaka, K. et al., Giant Rashba-type spin splitting in
bulk BiTeI. Nat. Mat. \textbf{10}, 521 (2011).

\bibitem{EuO} Matthias, B. T., Bozorth, R. N. \& Van Vleck, J. H. Ferromagnetic interaction in EuO. Phys. Rev. Lett. \textbf{7}, 160 (1961).


\end{thebibliography}
\end{document}